\documentclass[12pt]{article}
\usepackage{lscape}
\usepackage{citesort}
\usepackage{amssymb}
\usepackage{appendix}
\usepackage{multirow}

\begin{document}

\def\qq{\langle \bar q q \rangle}
\def\uu{\langle \bar u u \rangle}
\def\dd{\langle \bar d d \rangle}
\def\sp{\langle \bar s s \rangle}
\def\GG{\langle g_s^2 G^2 \rangle}
\def\Tr{\mbox{Tr}}
\def\figt#1#2#3{
        \begin{figure}
        $\left. \right.$
        \vspace*{-2cm}
        \begin{center}
        \includegraphics[width=10cm]{#1}
        \end{center}
        \vspace*{-0.2cm}
        \caption{#3}
        \label{#2}
        \end{figure}
	}
	
\def\figb#1#2#3{
        \begin{figure}
        $\left. \right.$
        \vspace*{-1cm}
        \begin{center}
        \includegraphics[width=10cm]{#1}
        \end{center}
        \vspace*{-0.2cm}
        \caption{#3}
        \label{#2}
        \end{figure}
                }

\def\ds{\displaystyle}
\def\beq{\begin{equation}}
\def\eeq{\end{equation}}
\def\bea{\begin{eqnarray}}
\def\eea{\end{eqnarray}}
\def\beeq{\begin{eqnarray}}
\def\eeeq{\end{eqnarray}}
\def\ve{\vert}
\def\vel{\left|}
\def\ver{\right|}
\def\nnb{\nonumber}
\def\ga{\left(}
\def\dr{\right)}
\def\aga{\left\{}
\def\adr{\right\}}
\def\lla{\left<}
\def\rra{\right>}
\def\rar{\rightarrow}
\def\lrar{\leftrightarrow}  
\def\nnb{\nonumber}
\def\la{\langle}
\def\ra{\rangle}
\def\ba{\begin{array}}
\def\ea{\end{array}}
\def\tr{\mbox{Tr}}
\def\ssp{{\Sigma^{*+}}}
\def\sso{{\Sigma^{*0}}}
\def\ssm{{\Sigma^{*-}}}
\def\xis0{{\Xi^{*0}}}
\def\xism{{\Xi^{*-}}}
\def\qs{\la \bar s s \ra}
\def\qu{\la \bar u u \ra}
\def\qd{\la \bar d d \ra}
\def\qq{\la \bar q q \ra}
\def\gGgG{\la g^2 G^2 \ra}
\def\GG{\langle g_s^2 G^2 \rangle}
\def\g5{\gamma_5 \not\!q}
\def\x{\gamma_5 \not\!x}
\def\g5{\gamma_5}
\def\sb{S_Q^{cf}}
\def\sd{S_d^{be}}
\def\su{S_u^{ad}}
\def\sbp{{S}_Q^{'cf}}
\def\sdp{{S}_d^{'be}}
\def\sup{{S}_u^{'ad}}
\def\ssp{{S}_s^{'??}}

\def\sig{\sigma_{\mu \nu} \gamma_5 p^\mu q^\nu}
\def\fo{f_0(\frac{s_0}{M^2})}
\def\ffi{f_1(\frac{s_0}{M^2})}
\def\fii{f_2(\frac{s_0}{M^2})}
\def\O{{\cal O}}
\def\sl{{\Sigma^0 \Lambda}}
\def\es{\!\!\! &=& \!\!\!}
\def\ap{\!\!\! &\approx& \!\!\!}
\def\ar{&+& \!\!\!}
\def\arrr{\!\!\!\! &+& \!\!\!}
\def\ek{&-& \!\!\!}
\def\vev{&\vert& \!\!\!}
\def\kek{\!\!\!\!&-& \!\!\!}
\def\cp{&\times& \!\!\!}
\def\se{\!\!\! &\simeq& \!\!\!}
\def\eqv{&\equiv& \!\!\!}
\def\kpm{&\pm& \!\!\!}
\def\kmp{&\mp& \!\!\!}
\def\mcdot{\!\cdot\!}
\def\erar{&\rightarrow&}


\def\simlt{\stackrel{<}{{}_\sim}}
\def\simgt{\stackrel{>}{{}_\sim}}

\title{
         {\Large
                 {\bf
Magnetic moment of $X_Q$ state with $J^{PC}=1^{+\pm}$ in light cone QCD sum
rules
                 }
         }
      }

\author{\vspace{1cm}\\
{
{\small A. K. Agamaliev$^1$\thanks{e-mail: guluoglu@mail.ru}}\,,
{\small T. M. Aliev$^2$\thanks{e-mail: taliev@metu.edu.tr}~\footnote{
permanent address:Institute of Physics, Baku, Azerbaijan}}\,\,,
{\small M. Savc{\i}$^2$\thanks{e-mail: savci@metu.edu.tr}}
} \\
{\small $^1$ Faculty of Physics, Baku State University, Az--1148,
Baku, Azerbaijan 
}\\
{\small $^2$ Physics Department, Middle East Technical University,
06531 Ankara, Turkey }}

\date{}

\begin{titlepage}
\maketitle
\thispagestyle{empty}

\begin{abstract}

The magnetic moments of the recently observed resonance $X_b(5568)$
by DO Collaboration and its
partner with charm quark are calculated in the framework of the light cone QCD
sum rules, by assuming that these resonances are represented as tetra--quark
states with quantum numbers $J^{PC}=1^{+\pm}$.
The magnetic moment can play critical role
in determination of the quantum numbers, as well as giving useful
information about the inner structure of these mesons.

\end{abstract}


\end{titlepage}

\section{Introduction}

Observation of the charmonium like $X(3872)$ resonance by the Belle
Collaboration \cite{Rdmn01} opens new directions in particle physics. At
present, more than 23 new exotic particles are discovered experimentally
(for a review, see \cite{Rdmn02}). The usual thing about these discoveries is
that, these states denoted as XYZ family, cannot be described by
the usual quark--antiquark picture, and these are indications that they have more
complicated structures. Experiments are usually focused on the measurement of
their masses, identifications of their spins, parities, as well as decay
widths.

These experimental observations have stimulated theoretical studies in this
direction. In theoretical analysis for studying the properties of exotic
states, usually two
pictures, tetra--quark or molecular picture (bound states of two
mesons), are used (for a review see \cite{Rdmn03,Rdmn04} and \cite{Rdmn05}).  
Recently the DO collaboration has reported the observation of the $X_b(5568)$ state
\cite{Rdmn06}, whose analysis assigned the quantum numbers $J^{PC}=0^{++}$,
while did not exclude the possibility $J^{PC}=1^{++}$.
Later however, LHCb \cite{Rdmn06} and CMS \cite{Rdmn07} Collaborations did
not confirm the existence of this state.
Therefore, the observed $X_b(5568)$ resonance by DO Collaboration
still needs solid experimental conformation.
The analysis of the mass, decay width $X_b \to B_s^0 \pi$ has already
been performed with the quantum number $J^{PC}=0^{++}$ in many works
(see \cite{Rdmn09}--\cite{Rdmn18}).    

In the present work, we assume that $X_b(5568)$ state  and its charm partner
(hereafter we will denote these states as $X_Q$) has quantum numbers
$J^{PC}=1^{+\pm}$, and calculate the magnetic moment of these exotic states.
Obviously, the $X_Q$ state with the quantum numbers $J^{PC}=0^{++}$
has zero magnetic moment.

The paper is organized as follows. In section 2, we consider $X_b(5568)$
state, and its partner with $c$--quark as a tetra--quark state with the
quantum numbers $J^{PC}=1^{+\pm}$, and
calculate the magnetic moment of these states in the framework of the light cone
QCD sum rules method. Section 3 is devoted to the numerical analysis of the
sum rules obtained for this exotic state.

\section{The $X_Q$ meson magnetic moment in light cone QCD sum rules} 

In calculating the magnetic moment of the $X_Q$ meson in the framework of the light
cone QCD sum rules, we start by considering the following correlation
function,
\bea
\label{edmn01}
\Pi_{\mu\nu\rho} (p,q) = - \int d^4y e^{ipx} \lla 0 \vel \mbox{T}
\left\{J_\mu^{X_Q}(x) J_\nu (y) J_\rho^{X_Q\dagger} (0) \right\}\ver 0 \rra\,.
\eea
Here, the current $J^{X_Q}$ with the quantum number $J^{PC}=1^{+\pm}$ describes
the corresponding $X_Q$ meson, and has the form,
\bea
\label{edmn02}
J_\mu^{X_Q} \es {1\over \sqrt{2}} \varepsilon^{abc}\varepsilon^{ade} 
\Big\{\left( s^{bT} (x) C \gamma_5 u^c \right)
\left( \bar{Q}^{d} (x) C \gamma_\mu \bar{d}^{eT} \right) \pm
\left( s^{bT} (x) C \gamma_\mu u^c \right)
\left( \bar{Q}^{d} (x) C \gamma_5 \bar{d}^{eT} \right) \Big\}\,,
\eea
where $a,b,c,d,e$ are the color indices, $C$ is the charge conjugation
operator. The electromagnetic current $J_\nu$ is given as,
\bea
J_\nu = e_s \bar{s} \gamma_\nu s + e_u \bar{u} \gamma_\nu u +e_d \bar{d}
\gamma_\nu d + e_Q \bar{b} \gamma_\nu b\,,\nnb
\eea
where $e_q$ is the corresponding electric charge.

From technical point of view, it is quite useful to rewrite the correlation
function by introducing the background electromagnetic plane wave field
\bea
F_{\mu\nu} = i \left( \varepsilon_\nu^\lambda q_\mu -
\varepsilon_\mu^\lambda q_\nu \right)\,,\nnb
\eea
where $\varepsilon_\nu$ and $q_\mu$ are the polarization and four--momentum
of the background electromagnetic field. The correlator can then be written
as,
\bea
\label{edmn03}
\Pi_{\mu\nu\rho} (p,q) \varepsilon^{(\lambda)\nu} = i \int d^4y e^{ipx} \lla 0
\vel \mbox{T} \left\{J_\mu^{X_Q}(x) \bar{J}_\rho^{X_Q} (0) \right\}\ver 0
\rra_F\,.
\eea
The subscript in this expression means that the expectation value is
evaluated in the background electromagnetic $F_{\mu\nu}$ field.
The correlation function (\ref{edmn01}) can be obtained by expanding
Eq. (\ref{edmn03}) in powers of the background field, and keeping only terms
linear in $F_{\mu\nu}$, which corresponds to the single photon emission
(see \cite{Rdmn19} and \cite{Rdmn20} for more details about
the background field).

We first calculate the correlation function in Eq. (\ref{edmn01})
in terms of the hadronic degrees
of freedom. To do that,
we saturate it with complete set of intermediate states
having the same quantum numbers as the interpolating current of $X_Q$, and
then isolate the contributions of the ground state, from which we get, 
\bea
\label{edmn04}
\Pi_{\mu\nu\rho} (p,q) \varepsilon^{(\lambda)\nu} = {\lla 0 \vel J_\mu^{X_Q} \ver
X_Q(p) \rra \over p^2 - m_{X_Q}^2} \lla X_Q(p) \Big\ve X_Q(p+q) \rra_F
{\lla X_Q(p+q)\vel \bar{J}_\rho^{X_Q} \ver 0 \rra \over
(p+q)^2 - m_{X_Q}^2} + \cdots
\eea
where the dots represent the contributions coming from the higher states and
continuum, and $q$ is the momentum of the photon. The matrix element
$\lla 0 \vel X_\mu^{X_Q} \ver X_Q \rra$ is determined as,
\bea
\label{edmn05}
\lla 0 \vel X_\mu^{X_Q} \ver X_Q \rra = \lambda_{X_Q}
\varepsilon_\mu^\lambda\,.
\eea

Using the time reversal and parity invariance, the electromagnetic vertex
of the two vector mesons in presence of the electromagnetic background field
can be defined in terms of the three form factors
as \cite{Rdmn21}:
\bea
\label{edmn06}
\lla X_Q (p,\varepsilon^\lambda) \ve  X_Q (p+q,\varepsilon^{\lambda^\prime})
\rra_F \es - \varepsilon^\tau (\varepsilon^{\lambda})^\alpha
(\varepsilon^{\lambda^\prime})^\beta
\Bigg\{ G_1(Q^2) g_{\alpha\beta} (2p+q)_\tau +
G_2(Q^2) (q_\alpha g_{\tau\beta} - q_\beta g_{\tau\alpha}) \nnb \\
\ek {1\over 2 m_{X_Q}^2} G_3(Q^2) q_\alpha q_\beta (2p+q)_\tau \Bigg\}\,,
\eea
where $\varepsilon^\tau$ is the photon, and $(\varepsilon^\lambda)^\alpha$,
$(\varepsilon^{\lambda^\prime})^\beta$ are the polarizations of $X_Q$
mesons. The form factors $G_i(Q^2)$ can be written in terms of charge
$F_C(Q^2)$,
magnetic $F_M(Q^2)$ and quadrapole $F_{\cal D}(Q^2)$ form factors in the following way. 
\bea
\label{edmn07}
F_C(Q^2) \es G_1(Q^2) + {2\over 3} \eta F_{\cal D}(Q^2)\,,\nnb \\
F_M(Q^2) \es G_2(Q^2)\,,\nnb \\
F_{\cal D}(Q^2) \es G_1(Q^2)-G_2(Q^2)+(1+\eta) G_3(Q^2)\,,
\eea
where $\eta=Q^2/4 m_{X_Q}^2$. At zero momentum--square transfer, the form
factors $F_C(Q^2=0)$, $F_M(Q^2=0)$, and $F_{\cal D}(Q^2=0)$ are related to the
electric charge, magnetic moment $\mu$ and the quadrapole moment ${\cal D}$ in
the following way:
\bea
\label{edmn08}
e F_C(0) \es e \,, \nnb\\
e F_M(0) \es 2 m_{X_Q} \mu \,, \nnb\\
e F_{\cal D} \es m_{X_Q}^2 {\cal D}\,.
\eea
Using Eqs. (\ref{edmn04}), (\ref{edmn05}) and (\ref{edmn06}), and performing
summation over polarization of the $X_Q$ meson, and imposing the condition
$q\!\cdot\!\varepsilon = 0$, at $Q^2=0$ the correlation function takes the
form,
\bea
\label{edmn09}
\varepsilon^\nu \Pi_{\mu\nu\rho} \es \lambda_{X_Q}^2 \varepsilon^\tau {1\over m_{X_Q}^2 - (p+q)^2}
{1\over m_{X_Q}^2 - p^2} \Bigg\{2 p_\tau F_C(0) \Bigg[g_{\mu\rho} - {p_\mu
q_\rho-p_\rho q_\mu \over m_{X_Q}^2 } \Bigg] \nnb \\
\ar F_M (0) \Bigg[q_\mu g_{\rho\tau} - q_\rho g_{\mu\tau} +
{1\over m_{X_Q}^2} p_\tau (p_\mu q_\rho - p_\rho q_\mu ) \Bigg]
- [F_C(0) + F_{\cal D}(0)] {p_\tau \over m_{X_Q}^2 } q_\mu
q_\rho \Bigg\}\,.
\eea

In order to determine the magnetic moment of $X_Q$ meson from
Eq. (\ref{edmn09}) we choose the structure
$(p\!\cdot\!\varepsilon)(p_\mu q_\nu-p_\nu q_\mu)$. The reason why we prefer
this structure is that it contains more powers of momentum, which
exhibits good convergence of the operator product expansion (OPE), and hence
leads to more reliable determination of the magnetic moment $\mu$.
In result, for the coefficient of the structure
$(p\!\cdot\!\varepsilon)(p_\mu q_\nu-p_\nu q_\mu)$ of the correlation
function from hadronic side we obtain,
\bea
\label{edmn10}    
\Pi = {\lambda_{X_Q}^2 \over m_{X_Q}^2} {1 \over (m_{X_Q}^2 - p^2)
[m_{X_Q}^2 - (p+q)^2]} \,\mu \,,
\eea
where $\mu$ is the magnetic moment of $X_Q$ meson in natural units, i.e., in
units of $e/2 m_{X_Q}$.

In constructing the sum rule for the magnetic moment of $X_Q$ meson, we
further need to calculate the correlation function in terms of quark and
gluon degrees of freedom.

Using the explicit form of the interpolating current given in Eq.
(\ref{edmn02}) we get the following expression for the correlation function
in terms of the relevant propagators as,
\bea
\label{edmn11}
\Pi_{\mu\rho} \varepsilon_\nu^\lambda \es {i\over 2} \epsilon^{abc}
\epsilon^{ade} \epsilon^{a^\prime b^\prime c^\prime} \epsilon^{a^\prime
d^\prime e^\prime} \int d^4x e^{ipx} \lla 0 \ve 
\mbox{Tr} \gamma_5 \widetilde{S}_s^{bb^\prime} (x) \Gamma_5 S_u^{cc^\prime} (x)
\mbox{Tr} \gamma_\mu \widetilde{S}_Q^{dd^\prime} (-x) \Gamma_\rho S_d^{ee^\prime} (-x)
\right. \nnb \\
\kpm \left. \mbox{Tr} \gamma_\mu \widetilde{S}_s^{bb^\prime} (x) \Gamma_5 S_u^{cc^\prime} (x)
\mbox{Tr} \gamma_5 \widetilde{S}_Q^{dd^\prime} (-x) \Gamma_\rho S_d^{ee^\prime} (-x)
\right. \nnb \\
\kpm \left.\mbox{Tr} \gamma_5 \widetilde{S}_s^{bb^\prime} (x) \Gamma_\mu S_u^{cc^\prime} (x)
\mbox{Tr} \gamma_\rho \widetilde{S}_Q^{dd^\prime} (-x) \Gamma_5 S_d^{ee^\prime} (-x)
\right. \nnb \\
\ar \left. \mbox{Tr} \gamma_\rho \widetilde{S}_s^{bb^\prime} (x) \Gamma_\mu S_u^{cc^\prime} (x)
\mbox{Tr} \gamma_5 \widetilde{S}_Q^{dd^\prime} (-x) \Gamma_5 S_d^{ee^\prime} (-x)
\ve 0 \rra_F\,,
\eea
where $\widetilde{S}= CS^TC$. It follows from this expression that in order
to calculate the correlation function from the QCD side, the expressions of the
light and heavy quark propagators are needed in presence of the external
field. The light cone expansion of the light quark propagator is calculated
in \cite{Rdmn22} whose expression in $x$--space is,
\bea
\label{edmn12}
S_q(x) \es {i \rlap/x\over 2\pi^2 x^4} - {m_q\over 4 \pi^2 x^2}
- {\qq \over 12} \Bigg(1 - i {m_q \over 4} \rlap/{x} \Bigg) - {\qq \over 192}
m_0^2 x^2 \Bigg(1 - i {m_q \over 6} \rlap/{x} \Bigg) \nnb \\
\ek {i g_s \over 16 \pi^2 x^2} \int_0^1 du \Bigg[\rlap/{x} \bar{u}
\sigma_{\alpha \beta} G^{\alpha \beta} (ux)
+ u \sigma_{\alpha \beta} \rlap/{x} G^{\alpha \beta} (ux) \nnb \\
\ek {1\over 2} i m_q x^2 \sigma_{\alpha \beta} G^{\alpha \beta} (ux)
\left( \ln {-x^2 \Lambda^2\over 4}  +
 2 \gamma_E \right) \Bigg]\,,
\eea
where $\Lambda$ is the cut--of energy separating the perturbative and
nonperturbative domains, $\gamma_E$ is the Euler constant, $m_q$ is the
light quark mass. The light cone expansion of the heavy quark operator in
the $x$--space is given as,
\bea
\label{edmn13}
S_Q(x) \es {m_Q^2 \over 4 \pi^2} \Bigg\{ {K_1(m_Q\sqrt{-x^2}) \over \sqrt{-x^2}} +
i {\rlap/{x} \over -x^2} K_2(m_Q\sqrt{-x^2}) \Bigg\} \nnb \\ 
\ek {g_s \over 16 \pi^2} \int_0^1 du
G_{\mu\nu}(ux) \left[ \left(\sigma^{\mu\nu} \rlap/x + \rlap/x
\sigma^{\mu\nu}\right) {K_1 (m_Q\sqrt{-x^2})\over \sqrt{-x^2}} +
2 \sigma^{\mu\nu} K_0(m_Q\sqrt{-x^2})\right] +\cdots\,,
\eea
where $K_i(m_Q\sqrt{-x^2})$ are the modified Bessel functions.
Substituting Eqs. (\ref{edmn12}) and (\ref{edmn13}) in Eq. (\ref{edmn11}),
one can calculate the correlation function from the QCD side. This correlation
function receives both perturbative, i.e., when photon interacts perturbatively
with quark propagators, and nonperturbative, i.e., photon interacts with
light quarks at large distance, contributions.  

It should be noted here that, in calculating the perturbative contributions
one of the free quark operators, i.e., the first two terms in Eqs. (\ref{edmn12})
and (\ref{edmn13}) are replaced by,
\bea
S^{free} \to \int d^4y\, S^{free} (x-y)\,\rlap/{\!A}(y)\, S^{free} (y)\,,\nnb
\eea
and the remaining three propagators are taken as the free ones. In calculation of
the nonperturbative contributions it is necessary to replace one of the light
quark propagators by,
\bea
\label{edmn14}
S_{\alpha\beta}^{ab} \to -{1\over 4} \left(\bar{q}^a \Gamma_i q^b \right)
(\left(\Gamma_i\right)_{\alpha\beta}\,,
\eea
where $\Gamma_i$ are the full set of Dirac matrices,
and the remaining quark propagators are taken as in Eqs.
(\ref{edmn12}) and (\ref{edmn13}). Once 
Eq. (\ref{edmn14}) is plugged in Eq. (\ref{edmn11}), there appear matrix
elements such as $\lla \gamma(q)\vel \bar{q}(x) \Gamma_i q(0) \ver 0\rra$
and $\lla \gamma(q)\vel \bar{q}(x) \Gamma_i G_{\alpha\beta}q(0) \ver 0\rra$,
which are needed in calculating the nonperturbative contributions. In
addition to these matrix elements, in principle, nonlocal operators such as
$\bar{q} G^2 q$ and $\bar{q}q\bar{q}q$ are expected to appear. But it is
known that the contributions of such operators are small, which is justified    
by the conformal spin expansion \cite{Rdmn22,Rdmn23}, and therefore we shall
neglect them.
The matrix elements $\lla \gamma(q)\vel \bar{q}(x) \Gamma_i q(0) \ver 0\rra$  
and $\lla \gamma(q)\vel \bar{q}(x) \Gamma_i G_{\alpha\beta}q(0) \ver 0\rra$
are expressed in terms of the photon distribution amplitudes whose expressions
are given below.

\bea
&&\langle \gamma(q) \vert  \bar q(x) \sigma_{\mu \nu} q(0) \vert  0
\rangle  = -i e_q \qq (\varepsilon_\mu q_\nu - \varepsilon_\nu
q_\mu) \int_0^1 du e^{i \bar u qx} \left(\chi \varphi_\gamma(u) +
\frac{x^2}{16} \mathbb{A}  (u) \right) \nnb \\ &&
-\frac{i}{2(qx)}  e_q \qq \left[x_\nu \left(\varepsilon_\mu - q_\mu
\frac{\varepsilon x}{qx}\right) - x_\mu \left(\varepsilon_\nu -
q_\nu \frac{\varepsilon x}{q x}\right) \right] \int_0^1 du e^{i \bar
u q x} h_\gamma(u)
\nnb \\
&&\langle \gamma(q) \vert  \bar q(x) \gamma_\mu q(0) \vert 0 \rangle
= e_q f_{3 \gamma} \left(\varepsilon_\mu - q_\mu \frac{\varepsilon
x}{q x} \right) \int_0^1 du e^{i \bar u q x} \psi^v(u)
\nnb \\
&&\langle \gamma(q) \vert \bar q(x) \gamma_\mu \gamma_5 q(0) \vert 0
\rangle  = - \frac{1}{4} e_q f_{3 \gamma} \epsilon_{\mu \nu \alpha
\beta } \varepsilon^\nu q^\alpha x^\beta \int_0^1 du e^{i \bar u q
x} \psi^a(u)
\nnb \\
&&\langle \gamma(q) | \bar q(x) g_s G_{\mu \nu} (v x) q(0) \vert 0
\rangle = -i e_q \qq \left(\varepsilon_\mu q_\nu - \varepsilon_\nu
q_\mu \right) \int {\cal D}\alpha_i e^{i (\alpha_{\bar q} + v
\alpha_g) q x} {\cal S}(\alpha_i)
\nnb \\
&&\langle \gamma(q) | \bar q(x) g_s \tilde G_{\mu \nu} i \gamma_5 (v
x) q(0) \vert 0 \rangle = -i e_q \qq \left(\varepsilon_\mu q_\nu -
\varepsilon_\nu q_\mu \right) \int {\cal D}\alpha_i e^{i
(\alpha_{\bar q} + v \alpha_g) q x} \tilde {\cal S}(\alpha_i)
\nnb \\
&&\langle \gamma(q) \vert \bar q(x) g_s \tilde G_{\mu \nu}(v x)
\gamma_\alpha \gamma_5 q(0) \vert 0 \rangle = e_q f_{3 \gamma}
q_\alpha (\varepsilon_\mu q_\nu - \varepsilon_\nu q_\mu) \int {\cal
D}\alpha_i e^{i (\alpha_{\bar q} + v \alpha_g) q x} {\cal
A}(\alpha_i)
\nnb \\
&&\langle \gamma(q) \vert \bar q(x) g_s G_{\mu \nu}(v x) i
\gamma_\alpha q(0) \vert 0 \rangle = e_q f_{3 \gamma} q_\alpha
(\varepsilon_\mu q_\nu - \varepsilon_\nu q_\mu) \int {\cal
D}\alpha_i e^{i (\alpha_{\bar q} + v \alpha_g) q x} {\cal
V}(\alpha_i) \nnb \\ && \langle \gamma(q) \vert \bar q(x)
\sigma_{\alpha \beta} g_s G_{\mu \nu}(v x) q(0) \vert 0 \rangle  =
e_q \qq \left\{
        \left[\left(\varepsilon_\mu - q_\mu \frac{\varepsilon x}{q x}\right)\left(g_{\alpha \nu} -
        \frac{1}{qx} (q_\alpha x_\nu + q_\nu x_\alpha)\right) \right. \right. q_\beta
\nnb \\ && -
         \left(\varepsilon_\mu - q_\mu \frac{\varepsilon x}{q x}\right)\left(g_{\beta \nu} -
        \frac{1}{qx} (q_\beta x_\nu + q_\nu x_\beta)\right) q_\alpha
\nnb \\ && -
         \left(\varepsilon_\nu - q_\nu \frac{\varepsilon x}{q x}\right)\left(g_{\alpha \mu} -
        \frac{1}{qx} (q_\alpha x_\mu + q_\mu x_\alpha)\right) q_\beta
\nnb \\ &&+
         \left. \left(\varepsilon_\nu - q_\nu \frac{\varepsilon x}{q.x}\right)\left( g_{\beta \mu} -
        \frac{1}{qx} (q_\beta x_\mu + q_\mu x_\beta)\right) q_\alpha \right]
   \int {\cal D}\alpha_i e^{i (\alpha_{\bar q} + v \alpha_g) qx} {\cal T}_1(\alpha_i)
\nnb \\ &&+
        \left[\left(\varepsilon_\alpha - q_\alpha \frac{\varepsilon x}{qx}\right)
        \left(g_{\mu \beta} - \frac{1}{qx}(q_\mu x_\beta + q_\beta x_\mu)\right) \right. q_\nu
\nnb \\ &&-
         \left(\varepsilon_\alpha - q_\alpha \frac{\varepsilon x}{qx}\right)
        \left(g_{\nu \beta} - \frac{1}{qx}(q_\nu x_\beta + q_\beta x_\nu)\right)  q_\mu
\nnb \\ && -
         \left(\varepsilon_\beta - q_\beta \frac{\varepsilon x}{qx}\right)
        \left(g_{\mu \alpha} - \frac{1}{qx}(q_\mu x_\alpha + q_\alpha x_\mu)\right) q_\nu
\nnb \\ &&+
         \left. \left(\varepsilon_\beta - q_\beta \frac{\varepsilon x}{qx}\right)
        \left(g_{\nu \alpha} - \frac{1}{qx}(q_\nu x_\alpha + q_\alpha x_\nu) \right) q_\mu
        \right]
    \int {\cal D} \alpha_i e^{i (\alpha_{\bar q} + v \alpha_g) qx} {\cal T}_2(\alpha_i)
\nnb \\ &&+
        \frac{1}{qx} (q_\mu x_\nu - q_\nu x_\mu)
        (\varepsilon_\alpha q_\beta - \varepsilon_\beta q_\alpha)
    \int {\cal D} \alpha_i e^{i (\alpha_{\bar q} + v \alpha_g) qx} {\cal T}_3(\alpha_i)
\nnb \\ &&+
        \left. \frac{1}{qx} (q_\alpha x_\beta - q_\beta x_\alpha)
        (\varepsilon_\mu q_\nu - \varepsilon_\nu q_\mu)
    \int {\cal D} \alpha_i e^{i (\alpha_{\bar q} + v \alpha_g) qx} {\cal T}_4(\alpha_i)
                        \right\}\,,\nnb
\eea
where $\varphi_\gamma(u)$ is the leading twist-2, $\psi^v(u)$,
$\psi^a(u)$, ${\cal A}$ and ${\cal V}$ are the twist-3, and
$h_\gamma(u)$, $\mathbb{A}$, ${\cal T}_i$ ($i=1,~2,~3,~4$) are the
twist-4 photon distribution amplitudes (DAs),
and $\chi$ is the magnetic susceptibility. The values of
the input parameters in photon DAs are calculated in \cite{Rdmn20}.
The measure ${\cal D} \alpha_i$ is defined as
\bea
\int {\cal D} \alpha_i = \int_0^1 d \alpha_{\bar q} \int_0^1 d
\alpha_q \int_0^1 d \alpha_g \delta(1-\alpha_{\bar
q}-\alpha_q-\alpha_g)\,.\nnb
\eea

In order to obtain the sum rules for the magnetic moment of
the $X_Q$, and its partner with $c$--quark
it is necessary to match the hadronic and QCD side representations of the
correlation function. Performing double Borel transformation over
the variables $-p^2$ and $-(p+q)^2$, which suppress the higher states and
continuum contributions, we finally get the sum rules for the magnetic moment
of $X_Q$ state. Note that Borel transformations are carried out with the
help of the formulas,
\bea
\label{edmn15}
{\cal B}\Bigg\{ {1\over (p^2-m_1^2)[(p+q)^2-m_2^2]}\Bigg\} \to
e^{-m_1^2/M_1^2 - m_2^2/M_2^2}\,,
\eea
in the hadronic part, and
\bea
\label{edmn16}{
\cal B}\Bigg\{ {1\over [m^2 -\bar{u}p^2 - u(p+q)^2]^\alpha} \Bigg\} \to
(M^2)^{(2-\alpha)} \delta(u-u_o) e^{-m^2/M^2}\,,
\eea
where we replace
\bea
M^2={M_1^2 M_2^2 \over
M_1^2+M_2^2}\,,\mbox{and}~u_0={M_1^2\over M_1^2+M_2^2}\,.\nnb
\eea
in the theoretical part.

Since the mass of the final and initial $X_Q$ mesons are equal to each other, we
set $M_1^2=M_2^2=2 M^2$, which leads to $u_0=1/2$. Substituting these
relations into the sum rules obtained for the magnetic moment, we get
\bea
\label{edmn17}
\mu = {e^{m_{X_Q}^2/M^2} m_{X_Q}^2 \over \lambda_{X_Q}^2}\Bigg\{ \Pi^B
\Bigg\}\,.
\eea
Explicit expression of $\Pi^B$ for the $1^{++}$ state is given in the Appendix.

We can easily see from this expression that, in order to calculate
magnetic moment residue $\lambda_{X_Q}$ is also needed. This residue can be
determined by considering the two--point correlation function,
\bea
\label{edmn18}
\Pi_{\mu\nu} = i \int d^4x e^{ipx} \lla 0 \vel \mbox{T} \{ J_\mu^{X_Q} (x)
\bar{J}_\nu^{X_Q} (0) \} \ver 0 \rra\,.
\eea
Saturating this correlation function with $X_Q$ mesons, the phenomenological
part of (\ref{edmn18}) can be written as,
\bea
\label{edmn19}
\Pi_{\mu\nu}^{phys} = {\lambda_{X_Q}^2 \over m_{X_Q}^2 - p^2} \Bigg( -g_{\mu\nu} +
{p_\mu p_\nu \over p^2 } \Bigg) + \cdots
\eea
Choosing the coefficient of the structure $g_{\mu\nu}$ which describes the
contribution of the pure $1^+$ state, and performing Borel transformation
over $-p^2$, the physical part of the correlation function can be  written
as
\bea
\label{edmn20}
\Pi_2^{phys\,B} = \lambda_{X_Q}^2 e^{-m_{X_Q}^2/M^2}\,.
\eea
To be able to construct sum rules for $\lambda_{X_Q}$, the correlation
function (\ref{edmn18}) must be calculated from the QCD side. The expansion of
the correlation function in terms of the heavy and light quark propagators is given
by Eq. (\ref{edmn11}) in the absence of background field $F$.

Substituting Eq. (\ref{edmn12}) and (\ref{edmn13}) into Eq. (\ref{edmn18}),
and choosing the coefficient of the structure $g_{\mu\nu}$, and performing
Fourier and Borel transformations, for the residue of $X_Q$ we get
\bea
\label{edmn21}
\lambda_{X_Q}^2 = e^{m_{X_Q}^2/M^2} \Pi_2^{(theor\,B)}\,,  
\eea
where subscript 2 stands for the two--point correlation function. The expression for
$\Pi_2^{(theor\,B)}$ is quite lengthy and not illuminating, and therefore we do not present it
here explicitly.
It should be noted here that, from the numerical analysis of the
sum rules for the masses of the $1^{++}$ and $1^{+-}$ $X_c$ states, we get
\bea
m_{X_c} \simeq (2.55 \pm 0.15)\,GeV\,.\nnb
\eea

\section{Numerical analysis}

The key ingredient of the light cone QCD sum rules for the magnetic moment is the photon
distribution amplitudes (DAs). The values of DAs are determined in
\cite{Rdmn20}, which we will use in our numerical calculations.

Sum rules for the magnetic moment contain, in addition to DAs, also
many input parameters, such as the
masses of the strange, charm and bottom quarks, values of the quark condensates.
In our numerical analysis we use, $\overline{m}_b(m_b) = (4.18\pm 0.03)\,GeV$,
$\overline{m}_c(m_c) = (1.275\pm 0.025)\,GeV$
(in $\overline{MS}$ scheme), $m_s(2\,GeV)=(95\pm 5)\,MeV$,
$\qq\ve_{1\,GeV}=(-0.24\pm0.01)^3\,GeV^3$ \cite{Rdmn24,Rdmn25},
and we use $\chi(1\,GeV)=-2.85\,GeV^{-2}$ for the
value of the magnetic susceptibility which is obtained in \cite{Rdmn26}.

The sum rules for the magnetic moment of $X_Q$ contain also two auxiliary parameters,
i.e., Borel mass parameter $M^2$ and the continuum threshold $s_0$.  
So in determination of the magnetic moment of $X_Q$ meson we should find
such regions of these auxiliary parameters so that the dependence of
magnetic moment on them would be minimal.

The working region of the Borel parameter is determined from the convergence of
the operator product expansion series and suppression of the contributions
of the higher states and the continuum. The upper bound of the Borel parameter
is fixed by requiring that the contribution of higher states and the
continuum constitutes less than 40\% of the contribution coming from the
perturbative part. The lower bound is fixed from the condition that higher
twist contributions are less than the leading twist contributions.
Having these restrictions imposed we obtain $5\,GeV^2 \le M^2 \le 8\,GeV^2$
for the $X_b(5568)$; and $2\,GeV^2 \le M^2 \le 4\,GeV^2$ for the $X_c$
mesons, respectively, for
the working regions of Borel mass--square parameter.

The working region of the continuum threshold $s_0$ is determined in the
following way. The difference $\sqrt{s_0}-m_{ground}$, where $m_{ground}$ is
the ground state mass, is the energy necessary to excite this particle to
its first excited state. The analysis in various sum rules predicts that
this difference varies in the domain from $0.3\,GeV$ to $0.8\,GeV$. In
further analysis we shall use $\sqrt{s_0}-m_{ground}=0.5\,GeV$.

As an example in Figs. 1 and 2 we present the dependence of the magnetic moment
$\mu$ of the $1^{++}$ and $1^{+-}$ $X_b(5568)$ states on $M^2$,
at several fixed values of the continuum threshold $s_0$. We observe from
the figures that the magnetic moment exhibits indeed good stability in the
working region of $M^2$.

The dependence of the magnetic moment $\mu$ on $s_0$,
at various fixed values of $M^2$ is also studied, and it is observed 
that the magnetic
moment $\mu$ is almost stable under variation of $s_0$ in its working
region. Our final results for the magnetic moments of these states are,
\bea
\mu = \left\{ \begin{array}{rl}
(0.22 \pm 0.07)\mu_N, &\mbox{for}~X_b~~1^{++}\\
(0.24 \pm 0.08)\mu_N, &\mbox{for}~X_b~~1^{+-}\\
(0.075 \pm 0.015)\mu_N, &\mbox{for}~X_c~~1^{++}\\
(0.10 \pm 0.02)\mu_N, &\mbox{for}~X_c~~1^{+-}\\
\end{array} \nnb \right. 
\eea
where the error in the result can be attributed to the variations in
$M^2$ and $s_0$, as well as to the uncertainties in the values of input
parameters. It follows from these results that the magnetic moment of
$X_b(5568)$ meson is large enough to be measured in future
experiments.  

In summary, in the present work we calculate the magnetic moment of exotic
$X_b(5568)$ state and its partner with $c$--quark state with the quantum
numbers $J^{PC}=1^{+\pm}$ in framework of the QCD sum rules method.
Measurement of the magnetic of $X_b(5568)$ meson in future experiments
can be very useful
in determining the quantum numbers, as well as understanding the internal
structure of these exotic states.

\newpage

\section*{Appendix: 
{\bf Expression of the invariant function $\Pi^B$} for the $1^{++}$ $X_Q$ state}

Coefficient of
${1\over 2}\left(p_\mu q_\nu - p_\nu q_\mu \right)
\varepsilon\!\cdot\!p$ structure.

\bea
\Pi^B \es
%
%
%
{e^{-m_Q^2/M^2} \over 331776 \pi^2 M^4}
f_{3\gamma} \GG m_0^2 m_Q \Big[47 \dd (e_s + e_u) - 2 e_u \sp + 2 e_s
\uu\Big] \psi^a(1 - u_0) \nnb \\
%
%
\ek {e^{-m_Q^2/M^2} \over 165888 \pi^2 m_Q M^2} f_{3\gamma} \GG m_0^2      
\Big[47 \dd (e_s + e_u) - 2 e_u \sp + 2 e_s \uu\Big] \psi^a(1 - u_0) \nnb \\
%
%
\ar {1\over 6144 \pi^4} m_0^2 m_Q M^2 \Big[2 \dd e_b + (e_b - 2 e_d) (\sp -
\uu)\Big] ({\cal I}_1 - 2 m_Q^2 {\cal I}_2 + m_Q^4 {\cal I}_3) \nnb \\
%
%
\ar {1\over 36864 \pi^4} f_{3\gamma} \GG m_Q^2 M^2 ({\cal I}_2 - m_Q^2 {\cal I}_3)  
  \Big[2 e_d i_2({\cal V},1) + (e_s + e_u) \widetilde{i}_3({\cal V},1)\Big] \nnb \\
%
%
\ek {1\over 1152 \pi^4} \dd \GG m_Q M^2 (3 {\cal I}_2 - 4 m_Q^2 {\cal I}_3) 
   \Big[e_s + e_u - 2 e_d \widetilde{j}(h_\gamma)\Big] \nnb \\
%
%
\ar {1\over 18432 \pi^4}
m_Q M^2 \Big\{6 m_0^2 \Big[(e_u \sp - e_s \uu) ({\cal I}_1 -
4 m_Q^2 {\cal I}_2 + 3 m_Q^4 {\cal I}_3) \nnb \\
\ek \dd (e_s + e_u) (24 {\cal I}_1 - 95 m_Q^2 {\cal I}_2 +
71 m_Q^4 {\cal I}_3)\Big] - 2 \dd e_d \GG (3 {\cal I}_2 -
4 m_Q^2 {\cal I}_3) i_1({\cal S},1) \nnb \\
\ek \GG (e_s \sp - e_u \uu) (3 {\cal I}_2 - 4 m_Q^2 {\cal I}_3) 
    \widetilde{i}_2(\widetilde{\cal S},1) \nnb \\
\ar 8  f_{3\gamma} m_Q ({\cal I}_2 
    - m_Q^2 {\cal I}_3)
    (e_s + e_u) (\GG + 96 \dd m_Q \pi^2) \psi^a(1 - u_0) -
     e_d \GG \psi^a(u_0)\Big\} \nnb \\
%
%
\ek {1\over 3072 \pi^6}
m_Q M^4 \Big\{12 \dd e_d m_Q^2 \pi^2 ({\cal I}_2 - 2 m_Q^2 {\cal I}_3 +
m_Q^4 {\cal I}_4) i_1({\cal S},1) \nnb \\
\ek 2 \pi^2 (e_s \sp - e_u \uu) ({\cal I}_1 - 6 m_Q^2 {\cal I}_2 +
9 m_Q^4 {\cal I}_3 - 4 m_Q^6 {\cal I}_4)
\widetilde{i}_2(\widetilde{\cal S},1) \nnb \\
\ek m_Q ({\cal I}_2 - 2 m_Q^2 {\cal I}_3 + m_Q^4 {\cal I}_4)
\Big[e_d \GG - (e_s + e_u) (\GG + 96 \dd m_Q \pi^2)\nnb \\
\ar 192 \dd e_d m_Q \pi^2 \widetilde{j}(h_\gamma)\Big]\Big\} \nnb \\
%
%
\ek {1\over 3072 \pi^4}
f_{3\gamma} m_Q^2 M^6 \Big\{6 e_d m_Q^2 ({\cal I}_3 - 2 m_Q^2 {\cal I}_4 +
m_Q^4 {\cal I}_5) i_2({\cal V},1) \nnb \\
\ar (e_s + e_u) ({\cal I}_2 - 3 m_Q^4 {\cal I}_4 + 2 m_Q^6
{\cal I}_5) \widetilde{i}_3({\cal V},1) \nnb \\
\ar 48 m_Q^2 ({\cal I}_3 - 2 m_Q^2 {\cal I}_4 +
m_Q^4 {\cal I}_5) \Big[(e_s + e_u) \psi^a(1 - u_0) 
- e_d \psi^a(u_0)\Big] \Big\} \nnb \\
%
%
\ek {1\over 512 \pi^6} 
m_Q^2 M^8 \Big[4 (e_d - e_s - e_u) m_Q^2 ({\cal I}_3 -
3 m_Q^2 {\cal I}_4 + 3 m_Q^4 {\cal I}_5 - m_Q^6 {\cal I}_6) \nnb \\
\ar e_b ({\cal I}_2 - 4 m_Q^2 {\cal I}_3 +
6 m_Q^4 {\cal I}_4 - 4 m_Q^6 {\cal I}_5 + m_Q^8 {\cal I}_6)\Big] \nnb \\
%
%
\ek {e^{-m_Q^2/M^2}\over 221184 m_Q \pi^4}
\Big\{\GG m_0^2 \Big[47 \dd (e_s + e_u) - 2 e_u \sp + 2 e_s \uu\Big]               
(1 - 3 m_Q^2 e^{m_Q^2/M^2} {\cal I}_2) \nnb \\
\ar 16  f_{3\gamma} \pi^2 \Big[6 m_0^2 m_Q^2 e^{m_Q^2/M^2}
(e_u \sp - e_s \uu) ({\cal I}_1 - 2 m_Q^2 {\cal I}_2) \nnb \\
\ek 3 \dd (e_s + e_u) m_0^2 m_Q^2 e^{m_Q^2/M^2} (48 {\cal I}_1-
95 m_Q^2 {\cal I}_2) \nnb \\
\ar 8 \dd (e_s + e_u) \GG (1 - 3 m_Q^2 e^{m_Q^2/M^2}
{\cal I}_2)\Big] \psi^a(1 - u_0)\Big\} \nnb \\  
\ar {1\over 2304 \pi^2}                   
e_d  f_{3\gamma} m_0^2 m_Q (\sp - \uu) ({\cal I}_1 - m_Q^2 {\cal I}_2)             
\psi^a(u_0)\,.\nnb
\eea

The functions $i_n~(n=1,2)$, $\widetilde{i}_n~(n=1,2)$, and $\widetilde{j}(f(u))$
are defined as:
\bea
\label{nolabel}
i_1(\phi,f(v)) \es \int {\cal D}\alpha_i \int_0^1 dv
\phi(\alpha_{\bar{q}},\alpha_q,\alpha_g) f(v) \delta(k-u_0)\,, \nnb \\
\widetilde{i}_1(\phi,f(v)) \es \int {\cal D}\alpha_i \int_0^1 dv
\phi(\alpha_{\bar{q}},\alpha_q,\alpha_g) f(v) \delta(\widetilde{k}-u_0)\,, \nnb \\
i_2(\phi,f(v)) \es \int {\cal D}\alpha_i \int_0^1 dv
\phi(\alpha_{\bar{q}},\alpha_q,\alpha_g) f(v) \delta^\prime(k-u_0)\,, \nnb \\
\widetilde{i}_2(\phi,f(v)) \es \int {\cal D}\alpha_i \int_0^1 dv
\phi(\alpha_{\bar{q}},\alpha_q,\alpha_g) f(v) \delta^\prime(\widetilde{k}-u_0)\,, \nnb \\
\widetilde{j}(f(u)) \es \int_{u_0}^1 du (u-u_0) f(u)\,, \nnb \\
{\cal I}_n \es \int_{m_Q^2}^{s_0} ds\, {e^{-s/M^2} \over s^n}\,,\nnb \\
\eea

where 
\bea
k = \alpha_q + \alpha_g (1-v)\,,~~~~~
\widetilde{k} = \alpha_{\bar{q}} + \alpha_g v\,.\nnb
\eea

\newpage

\section*{Figure captions}
{\bf Fig. (1)} The dependence of the magnetic moment of
$X_b(5568)~1^{++}$ state
on the Borel parameter $M^2$, at several fixed values of the
continuum threshold $s_0$.\\\\
{\bf Fig. (2)} The same as Fig. (1), but for the
$X_b(5568)~1^{+-}$ state.

\newpage

\begin{figure}
\vskip 3. cm
    \includegraphics{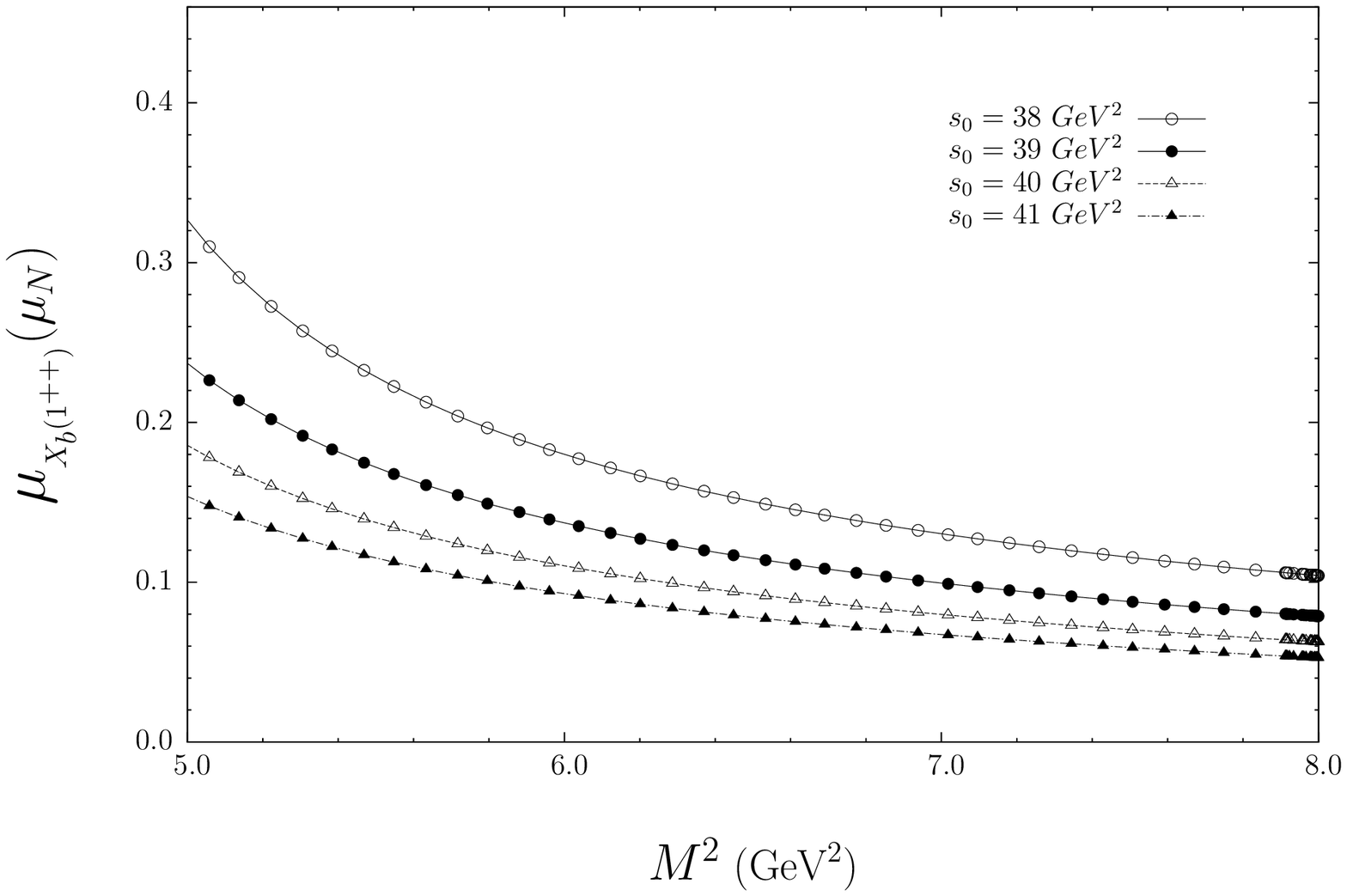}
\vskip 7.0cm
\caption{}
\end{figure}

\begin{figure}
\vskip 3. cm
    \includegraphics{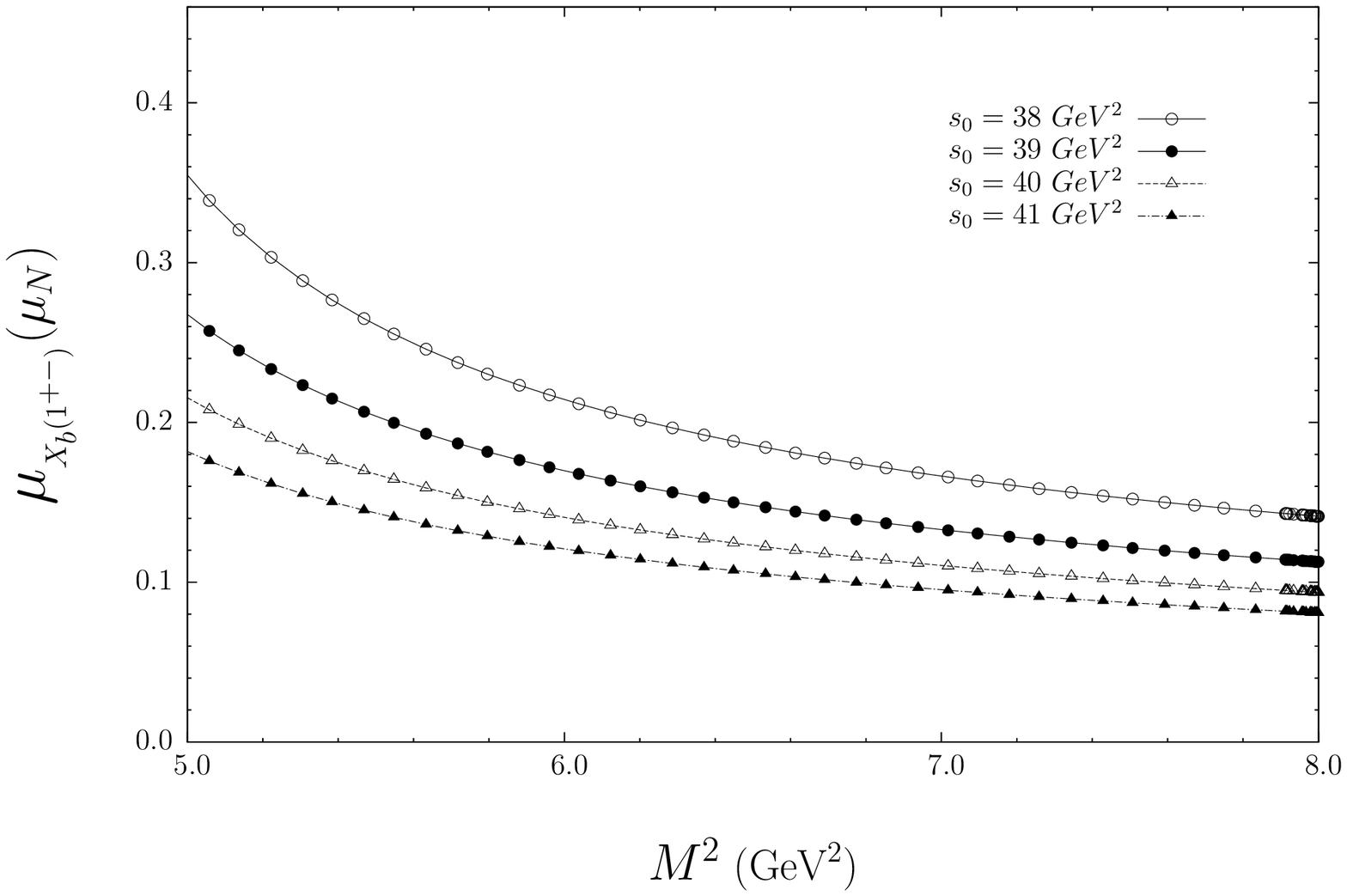}
\vskip 7.0cm
\caption{}
\end{figure}

\end{document}